\documentclass[column,showpacs,amsmath,amssymb,amsfonts,nofootinbib,plb]{revtex4}
\usepackage{graphicx}
\usepackage{dcolumn}
\usepackage{bm}
\usepackage{ulem}
\usepackage{overpic}
\usepackage{amssymb}

\newcommand{\nn}{\nonumber}

\def\beq{\begin{equation}}
\def\eeq{\end{equation}}

\usepackage{soul}
\usepackage{cancel}

\usepackage[usenames]{color}

\begin{document}
\input{epsf}

\title{On the solution of linearized (linear in $S$-matrix) Balitsky-Kovchegov equation}

%

\author{Raktim \surname{Abir}} 
\email{raktimabir@gmail.com, raktim.ph@amu.ac.in}

\author{Mariyah Siddiqah} 
\email{shah.siddiqah@gmail.com}

\affiliation{
Department of Physics, Aligarh Muslim University, Aligarh - $202002$, UP, India.\\
}

 \begin{abstract}
  We revisited solution of a linearized form of leading order Balitsky-Kovchegov equation
  (linear in $S$-matrix for dipole-nucleus scattering). Here we adopted 
 dipole transverse width dependent cutoff in order to regulate the dipole integral. We also have taken care of all the higher order 
 terms (higher order in the cutoff) that have been reasonably neglected before. The solution
 reproduces both McLerran-Venugopalan type initial condition (Gaussian in scaling variable) and Levin-Tuchin solution 
 (Gaussian in logarithm of scaling variable) in the appropriate limits. It also connects 
 this two opposite limits smoothly with better accuracy for sets of rescaled rapidity when compared to numerical solutions of full
 leading order Balitsky-Kovchegov equation.
 \end{abstract}


\date{\today}
\maketitle

 \section{Introduction}
 A typical scattering event in any high energy collider experiment usually involve rapidly growing 
 cascade of gluons. This is partly because high energy (and/or) high virtuality emitted gluons themselves emit 
 further gluons. At high enough energy this
 cascade of gluons may occupy all the available final state phase space to such an extent that fusion of multiple 
 gluons to single gluon begin to start. This could eventually develop a thermodynamical detail balance with 
 the multiple gluons produced from single gluon which leads to the origin of gluon saturation with a characteristic
 momentum scale $Q_s$ \cite{Gribov:1984tu}. This is a dynamically generated and energy
 dependent scale below which stochastic (almost) independent multiple scattering approximations are
 no longer valid and highly correlated non-linear gluon interactions dominates the phase space. 
 This gluon recombination also restores unitarity of the scattering $S$-matrix which will otherwise 
 violated by an exponential growth of gluon multiplicity. Consequently this saturation of gluons
 also avoids possible violation of Froissart bound 
 for the total scattering cross section through the power law growth of the Balistky-Fadin-Kuraev-Lipatov (BFKL) \cite{bfkl} solution which 
 encode energy evolution of the cross-section away from the non-linear region.  
 Unitary corrections to the BFKL equation in the Regge kinematics were first studied by Balitsky \cite{Balitsky:1995ub} within a Wilson 
 line formalism \cite{Balitsky:2001gj} and soon after by Kovchegov \cite{Kovchegov:1999yj,Kovchegov:1999ua} in the Muller's color dipole approach 
 \cite{Mueller:1993rr,Mueller:1994jq,Chen:1995pa}. 
 The Balistky hierarchic chain formed by the Wilson line operators reduced to 
 the closed form equation derived by Kovchegov in the large $N_c$ limit. 
 Integral kernel in the Balitsky-Kovchegov (BK) equation 
 for both linear 
 and non linear terms are identical and has a simple interpretation of splitting of one parent color 
 dipole into two daughter dipoles. 
 A lot of progress have been made since then in various aspects including solving the 
 equation both analytically and numerically and extending the equation beyond its leading order accuracy \cite{Kovchegov:2012mbw}. 
 %
 Next to leading order BK equation was derived \cite{Balitsky:2008zza},  
 inclusion of running coupling corrections to the BK evolution equations was done 
 \cite{Kovchegov:2006vj,Balitsky:2006wa,Albacete:2004gw}. 
 Solution of the NLO BFKL equation has been found analytically 
 \cite{Chirilli:2013kca}. Application of the leading order equation extended to jet quenching studies \cite{Abir:2015qva}. 
 First numerical study for the solution to the NLO Balitsky-Kovchegov equation in coordinate 
 space has been performed recently \cite{Lappi:2015fma}. Large double logarithms resumed in the QCD evolution of 
 color dipoles \cite{Iancu:2015vea} and in accordance with the HERA data \cite{Iancu:2015joa}.
 An analytic BK solution based on the eigenfunctions of the truncated BFKL equation have been proposed recently 
 that reproduces the initial condition and the high energy 
 asymptotics of the scattering amplitude \cite{Bondarenko:2015fca}.
 \\
 
 In order to have the evolved solution of BK equation one usually starts with the initial condition for the evolution from 
 McLerran-Venugopalan model \cite{McLerran:1993ni,McLerran:1993ka,McLerran:1994vd}
 or from phenomenological Golec-Biernat and M.~Wusthoff model
  \cite{GolecBiernat:1999qd,GolecBiernat:1998js}. 
 The imaginary part
 of the dipole-nucleus amplitude for deep inelastic scattering of the dipole with a large nucleus takes the following form, 
 \begin{eqnarray} 
 N_{\rm MV}(x_{10}, Y)=1-S_{MV}(x_{10},Y)=1-\exp\left(-\kappa ~ x_{10}^2Q_s^2(Y)\right)~,
 \label{MV}
 \end{eqnarray}
 where $\kappa=1/4$ or could be fixed from the definition of the saturation scale $Q_s$ and $x_{10}$ being transverse width 
 of the parent dipole. 
 Eq.\eqref{MV} is taken as the initial condition for the evolution, and expected to be valid for some initial rapidity both inside and 
 outside the saturation region. 
 The $S$-matrix in Eq.\eqref{MV} is Gaussian in the scaling variable $\tau$ ($\tau=x_\perp Q_s(Y)$) 
 with a (model dependent) variance $1/\sqrt{\kappa}$. However 
 in ultra high energy limit where Levin-Tuchin solution \cite{Levin:1999mw} of Balitsky-Kovchegov equation is valid, $S$-matrix 
 has the following asymptotic expression, 
 \begin{eqnarray} 
 N_{\rm LT}(x_{10}, Y)=1-S_{\rm LT}(x_{10},Y)=
 1-\exp\left(-\frac{1+2i\nu_0}{4\chi\left(0,\nu_0\right)}\ln^2\left[x_{10}^2Q_s^2(Y)\right]\right)~,
 \label{LT}
 \end{eqnarray}
 Unlike Eq.\eqref{MV} $S$-matrix in Eq.\eqref{LT} is a Gaussian in $\ln \tau$ (not in $\tau$).  
 Solution that span over full kinematic range of saturation dynamics is expected to be in accordance with both the 
 McLerran-Venugopalan type initial condition and 
 the Levin-Tuchin solution  in their appropriate limits.
 In this article we have revisited the solution for  
 linearized  LO BK equation. By linearized we mean linear in $S$ (unlike BFKL which is linear in $N$)
 where the term quadratic in $S$ has not been taken.  
 With a modified $x_{\perp}$ dependent form of cutoff in the dipole integral we obtain the general solution as, 
 \begin{eqnarray}
 N_{\rm }(x_{10}, Y)=1-S_{\rm }(x_{10},y)=1-\exp\left(\frac{1+2i\nu_0}{2\chi\left(0,\nu_0\right)} 
 ~{\rm Li}_2\left[-\lambda_1 x_{10}^2Q_s^2(Y)\right]\right)~,
 \label{AS}
 \end{eqnarray}
 where ${\rm Li}_2$ is dilogarithm function and $\lambda_1(\approx 7.22)$ is a parameter which is be fixed by the definition of
 $Q_s$.
 Interestingly, Eq.\eqref{AS} as solution of the linearized BK equation 
 reproduces both Eq.\eqref{MV} (Gaussian in $\tau=x_{10}Q_s(Y)$) and Eq.\eqref{LT} (Gaussian in logarithm of $\tau = x_{10}Q_s(Y)$
 $i.e$ in $\ln \tau$) in the 
 limits $\tau=x_{10}Q_s(Y) \ll 1$ and $\tau=x_{10}Q_s(Y) \gg 1$
 respectively. It also connects 
 this two opposite limit smoothly with a better accuracy when compared to numerical solutions of full LO BK equation.

 \section{The dipole integral}
 One convenient way to address high energy scatterings in QCD is to express the problem in hand in terms of color dipoles 
 degrees of freedom. 
 This approach, originally proposed by Mueller \cite{Mueller:1993rr,Mueller:1994jq,Chen:1995pa}, is formulated in the 
 transverse coordinate space. It has the added advantage that transverse coordinates of the dipoles are not
 changed during rapidity (or energy) evolution. This makes it easier to include the saturation effects into the model. 
  Typically one starts with a quark (anti-quark) pair in order to calculate probability of emission of a soft gluon off this pair. 
 Both the quark and anti-quark are to follow light cone trajectories and emitted gluons are calculated in the eikonal approximations
 (the projectile do not suffers any recoil).
 Adding contributions  coming from the quark and anti-quark together with their interference 
 one gluon part of onium wave function found to be proportional to following integral kernel convoluted over the onium wave function 
 with no soft gluon \cite{Mueller:1993rr},   
 \noindent 
 \begin{eqnarray} 
 I_{\rm dip}
 \equiv \int d^2{x_{2}}~\frac{x_{10}^2}{x_{20}^2x_{21}^2} 
 \equiv \int \frac{(x-y)^2}{(x-z)^2(z-y)^2}d^2z ~.
 \label{Monica1}
 \end{eqnarray}
 Above kernel (together with a Sudakov type form factor)
 can be interpreted as emission probability  of a soft gluon 
 from the dipole with two pole located at $x$ and $y$. In the large $N_c$ limit the emitted gluon can be seen as quark (anti-quark) pair 
 and above formula can be interpreted  as probability of decay of original parent dipole at $(x,y)$ of transverse size $x_{10}\equiv |x-y|$  
 into two new daughter dipoles at $(x,z)$ and at $(z,y)$ with sizes $x_{20}\equiv |x-z|$ and $x_{21}\equiv |z-y|$. 
 In this section we will revisit derivation of above integral which is central to the dipole studies.
 The integral $I_{\rm dip}$ supplemented with the factor $\bar{\alpha}_s/2\pi $ could be interpreted as the differential probability of
 decay of one parent dipole of  transverse size $x_{10}(\equiv x_1-x_0)$
 into two daughter dipole of arbitrary  sizes.
 %
 Having note that $d^2x_{2}$ is equal to \cite{Mueller:1993rr},
 \begin{eqnarray}
 2\pi x_{02} x_{12} \int_{0}^{\infty} dk  k  J_0(kx_{10})  J_0(kx_{20})  J_0(kx_{21}) dx_{20}  dx_{12} 
 \end{eqnarray} 
 where $J_0(z)$ is Bessel function of the first kind, we now write Eq.\eqref{Monica1} as \cite{Kovchegov:2012mbw}, 
 \begin{eqnarray}
 I_{\rm dip} &=& 2\pi x_{10}^2 \int_{0}^{\infty} dk k J_0(k x_{10})  
                            \int_{0}^{\infty} \frac{dx_{20}}{x_{20}}  J_0(kx_{20}) 
                             \int_{0}^{\infty} \frac{dx_{21}}{x_{21}}  J_0(kx_{21}) ~. \label{Madhuri}
 \end{eqnarray}
 The integral in Eq.\eqref{Madhuri} over $x_{20}$ and $x_{21}$ is ill-defined until one specifies a way 
 to regulate the ultra-violate singularities at $x_{20},x_{21} \rightarrow 0$. 
 %
 In the dipole model studies on usually introduce lower cutoff $\rho$ into the $x_{20}$ and $x_{21}$ integrals. 
 This procedure was first adopted by Mueller in \cite{Mueller:1993rr} and followed in subsequent other studies \cite{Kovchegov:1999ua},
  \begin{eqnarray} 
  && \int_{0}^{\infty} \frac{dx}{x} J_{0}(kx)~ {\Rightarrow}\int_{\rho}^{\infty} \frac{dx}{x} J_{0}(kx) 
  = \ln \frac{2}{k\rho} - \gamma  + {\cal O}(\rho)  ~.\label{pompi}
 \end{eqnarray}
 Using $\rho$ as a cutoff as usually done in the dipole model studies is one way to regulate the integral. 
 Alternatively, for example, one could replace  $1/x^2 \rightarrow 1/(x^2 + \rho^2)$ for $x_{20}^2$ and $x_{21}^2$ in 
 the denominator in Eq.\eqref{Madhuri} which gives order zero modified Bessel function of second kind 
 $K_0(\rho k)$. There are other ways to regulate the integral. All these regularizations should give the same leading-order 
 result as $\rho \rightarrow 0$, but the sub-leading terms would depend on the regularization procedure that actually
 been followed. Inside the saturation region $\rho$ usually identified with  inverse saturation momentum $1/Q_s$. 
 In this study we have revisited this issue with following two points, \\
 
 (1) We have considered ${\cal O}(\rho)$ and all other higher order terms, in Eq.\eqref{pompi}, that have been 
     ignored earlier, 
  \begin{eqnarray} 
  \int_{\rho}^{\infty} \frac{dx}{x} J_{0}(kx)  
  &=& \ln \frac{2}{k\rho} - \gamma  + \frac{k^2\rho^2}{8} ~{}_2F_3\left(1, 1; ~2,2,2;~-\frac{1}{4}k^2\rho^2\right)~,\nn \\
  &=& \ln \frac{2}{k\rho} - \gamma+ \sum_{m=1}^{\infty}\frac{(-1)^{m+1}}{2^{2m}(m!)^2} \frac{1}{2m}k^{2m}\rho^{2m} ~. \label{Alia}
  \end{eqnarray}
  Simple ratio test confirms radius of convergence of this series is infinity. We derived compact closed form expression of
  $I_{\rm dip}$ that contains contributions from ${\cal O}(\rho)$ and all other higher order terms.  \\
  
 (2) In earlier studies the cutoff $\rho$ usually identified with inverse saturation momentum $1/Q_s$ as 
  \begin{eqnarray}
  x_{20},x_{21}  \geqslant \rho = \frac{1}{Q_s}  \label{AngelinaJolie1}
  \end{eqnarray} 
  or, equivalently,  
  \begin{eqnarray}
  \frac{1}{x_{20}^2},\frac{1}{x_{21}^2} \leqslant Q_s^2 = \frac{1}{\rho^2}  \label{AngelinaJolie2}
  \end{eqnarray}  
    In this study we have adopted similar regularization procedure as done earlier but 
    assumed a general $x_{10}$ dependent form of cutoff as,     
  \begin{eqnarray}
   x_{20},x_{21}  \geqslant \rho=\frac{1}{Q_s}\frac{1}{\sqrt{\lambda_1+\lambda_2 (1/x_{10}Q_s)^2}} \label{AngelinaJolie3}
  \end{eqnarray}
   Eq.\eqref{AngelinaJolie3} actually implies, 
  \begin{eqnarray}
  \frac{1}{x_{20}^2},\frac{1}{x_{21}^2} \leqslant \lambda_1 Q_s^2 + \frac{\lambda_2}{x_{10}^2}= \frac{1}{\rho^2}  \label{AngelinaJolie4}
  \end{eqnarray} 
  which can be compared with Eq.\eqref{AngelinaJolie2}. 
  Here $\lambda_1$ and $\lambda_2$ are two positive real parameter which would be fixed in the following way, \\ 
  
  (2.a) Parameter $\lambda_2$ would be fixed by requiring the fact that in the limit
  $x_{10}\rightarrow 0$ the dipole integral in Eq.\eqref{Madhuri} vanishes $i.e.$ $I_{\rm dip}\rightarrow 0$. \\
  
  (2.b) Parameter $\lambda_1$ would be fixed by the definition of saturation momentum :  at $x_{10}=1/Q_s$, 
  numerical value of $S$-matrix would be half, 
  \begin{eqnarray}
  S\left(x_{10}=1/Q_s\right)=\frac{1}{2}~. 
  \end{eqnarray}  
  \noindent 
  Eq. \eqref{AngelinaJolie4} is just an ad hoc ansatze for the UV cutoff as the generalization of Eq.(10). 
  However this modified form of cutoff ensures that in both the limit, $Q_s \rightarrow \infty ~(x_{10}~ {\rm fixed})$ and
  $x_{10} \rightarrow 0 ~(Q_s~ {\rm fixed})$,
  the cutoff tends to zero, $\rho \rightarrow 0$. Hence, unlike earlier studies where result is valid only in the limit
  $Q_s\rightarrow \infty$, here the final result is expected to be
  valid and free from regularization scheme artifacts both in the limit $x_{10}Q_s \gg 1$ as well as 
  $x_{10}Q_s \ll 1$. With above mentioned modifications $I_{\rm dip}$ found to be, (details are in Appendix I),
  \begin{eqnarray}
  I_{\rm dip} &=& 2\pi \ln\left(\lambda_1 x_{10}^2Q_s^2+\lambda_2\right) 
  -2\pi {\rm Li}_{1}\left(\frac{1}{\lambda_1 x_{10}^2Q_s^2+\lambda_2}\right)  ~.
              \label{Moutusi}
  \end{eqnarray}
 Having note that ${\rm Li}_1(z)=-\ln(1-z)$ one may further simplify as,  
  \begin{eqnarray}
  I_{\rm dip} &=& 2\pi \ln\left(\lambda_1 x_{10}^2 Q_s^2+\lambda_2-1\right) ~, \label{DewBarrymore1} \\
              &=& 2\pi \ln\left(\lambda_1 x_{10}^2Q_s^2\right) +2\pi \ln\left(1-\frac{1-\lambda_2}{\lambda_1 x_{10}^2Q_s^2}\right)  ~, \nn \\
              &=& 2\pi \ln\left(\lambda_1 x_{10}^2 Q_s^2\right)- 
              2\pi {\rm Li}_{1}\left(\frac{1-\lambda_2}{\lambda_1 x_{10}^2Q_s^2}\right)~.\label{DewBarrymore2}  
  \end{eqnarray}
  Looking at Eq.\eqref{DewBarrymore1} one could fix $\lambda_2$ as $\lambda_2=2$ by taking the limit 
  $I_{\rm dip}\rightarrow0$ when $x_{10}\rightarrow0$. Therefore, we have, 
  \begin{eqnarray}
  I_{\rm dip} &=& 2\pi \ln\left(\lambda_1 x_{10}^2 Q_s^2+1\right)~. 
  \end{eqnarray}
  The parameter $\lambda_1$ would be fixed from the definition of $Q_s$, in the next section, as mentioned earlier. 
  %

  

  
 \section{S-Matrix inside saturation region}
  \begin{figure}
  \includegraphics[width=0.6\linewidth]{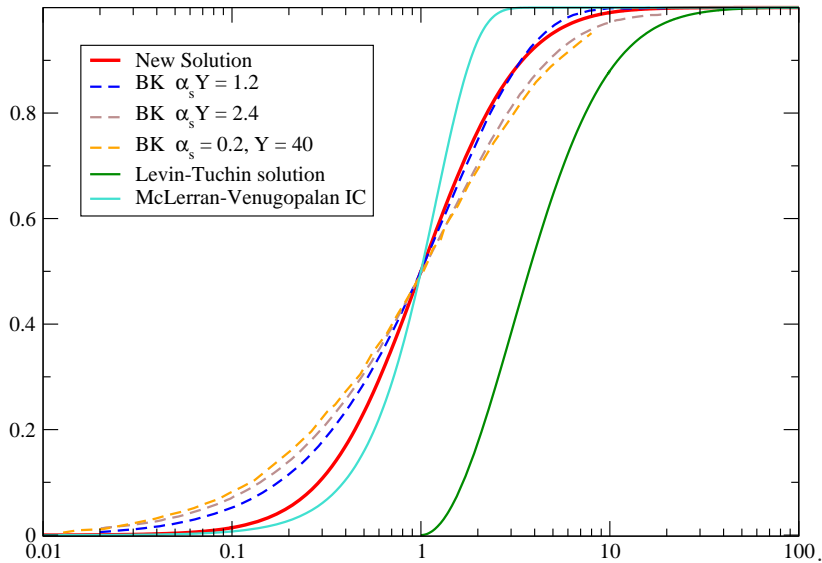}\label{N_vs_xQ}.
 %
  \caption{The dipole amplitude $N(x_{10},Y)$ as function of scaling variable $\tau=x_{10}Q_s(Y)$;
 new solution (Eq.\eqref{AS}) compared with numerical solutions of leading order 
 Balitsky-Kovchegove equation for two sets of rescaled rapidity, $\alpha_s Y=1.2,2.4$, one set of fixed coupling $\alpha_{s}=0.2$
 for $Y = 40$ \cite{Albacete:2007yr},    
 McLerran-Venugopalan initial
 condition Eq.\eqref{MV} (tweaked to reproduce $N=1/2$ at $\tau=1$) and Levin-Tuchin solution Eq.\eqref{LT} also displayed for reference.} 
 \label{SexyFigure1}
 \end{figure}
 \noindent Scattering $S$-matrix for the color dipole interacting with a large 
 nuclear target can be expressed as expectation value of two light-like path ordered Wilson lines 
 transversely separated by $x_{10}(={\bf x_\perp-y_\perp})$ as, 
 \begin{eqnarray}
 S(x_{10},Y)= \frac{1}{N_c}\langle {\rm Tr}\left[W(x_\perp,Y) W^{\dagger}(y_\perp,Y)\right]\rangle ~.
 \end{eqnarray}
 In large $N_c$ limit non-linear energy (or rapidity) evolution of the $S$-matrix is governed by 
 Balitsky-Kovchegov equation, 
 \begin{eqnarray}
 \frac{\partial}{\partial Y} S(x_{01}, Y) &&= \frac{\alpha_s N_c}{2\pi^2} 
                             \int d^2{x_{2}}~\frac{x_{01}^2}{x_{02}^2x_{21}^2}  
                             \left[S(x_{02}, Y) S(x_{12}, Y) - S(x_{01}, Y) \right]~.
 \label{Monalisa}
 \end{eqnarray}
 %
 Within the kinematic  
 domain where $S(x_{10}) \gg S(x_{02}, Y) S(x_{12}, Y)$, one can neglect the term quadratic in $S$ in Eq.\eqref{Monalisa}
 and the BK equation, an integro-differential equation in general, becomes first order partial differential
 equation of $S(x_{01}, Y)$, 
 \begin{eqnarray}
 \frac{\partial}{\partial Y} S(x_{01}, Y) 
 &=& -\frac{\alpha_s N_c}{2\pi^2} \int d^2{x_{2}}~\frac{x_{01}^2}{x_{02}^2x_{21}^2}~ S(x_{01}, Y) \label{Priyanka1} ~,  
 \end{eqnarray}
 In general one expect validity of this linear equation in the limit when $x_{10}>x_{20}$, $x_{21}>1/Q_s(Y)$.
 Here we note that when $x_{20}$, $x_{21} > x_{10}$  we could also expect $S(x_{20})S(x_{21}) < S(x_{10})$ $i.e.$
       quadratic term is smaller than the linear term in BK equation. Therefore this linearized form should be expected to
       valid (at least approximately) in both the limiting domain defined by $x_{10} > x_{20}, x_{21}$ and  $x_{10} < x_{20}, x_{21}$. 
 In Eq.\eqref{Priyanka1} the integral over dipole size goes over $x_{02},x_{12}> \rho$ as discussed in the Sec. II with, 
 \begin{eqnarray} 
 \rho=  \frac{1}{Q_s}\frac{1}{\sqrt{\lambda_1+2/(Q_s^2x_{10}^2)}}~,
 \end{eqnarray} 
 where $\lambda_2$ have already been fixed at 2. Using Eq.\eqref{DewBarrymore2}, Eq.\eqref{Priyanka1} can now be written as, 
 \begin{eqnarray}
 \frac{\partial}{\partial Y} \ln S(x_{01}, Y)  
 &=& -\bar{\alpha}_s \ln\left[\lambda_1 x_{10}^2Q_s^2(Y)\right] -\bar{\alpha}_s{\rm Li}_{1}\left(-\frac{1}{\lambda_1x_{10}^2Q_s^2(Y)}\right)~.  \label{Suchitra}
 \end{eqnarray}

 \noindent Solution of Eq.\eqref{Suchitra} can be written straightforwardly as,
 \begin{eqnarray}
 S=S_{0}\exp\left(\frac{1+2i\nu_0}{2\chi\left(0,\nu_0\right)}\left[\frac{1}{2}\ln^2\left(\lambda_1 x_{10}^2 Q^2_s(Y)\right)
 +{\rm Li}_2\left(-\frac{1}{\lambda_1 x_{10}^2 Q^2_s(Y)}\right)\right]\right) ~.  \label{DemiMoore}
 \end{eqnarray}
 Here we have used following leading order expression for saturation momentum \cite{Gribov:1984tu},  
 \begin{eqnarray}
 Q_s(Y)=Q_{s0}\exp\left(\bar{\alpha}_s\frac{\chi\left(0,\nu_0\right)}{1+2i\nu_0}Y\right)
 \approx Q_{s0}~e^{2.44\bar{\alpha}_sY}  \label{Mamta}
 \end{eqnarray}
 where,
 \begin{eqnarray}
 \chi(0,\nu)=2\psi(1)-\psi\left(\frac{1}{2}+i\nu\right)-\psi\left(\frac{1}{2}-i\nu\right)
 \end{eqnarray}
 and $\psi$ being digamma function and $S_{0}$ is a constant independent of any initial condition.
 On can further simplify Eq.\eqref{DemiMoore} to,
 \begin{eqnarray}
 S=S_{0}\exp\left(\frac{1+2i\nu_0}{2\chi\left(0,\nu_0\right)}
 \left[{\rm Li}_2\left(-\lambda_1 x_{10}^2 Q^2_s(Y)\right)\right]\right)~,   \label{PriyankaChopra}
 \end{eqnarray}
 where we have used following identity of dilogarithm for $x>0$,
 \begin{eqnarray} 
 {\rm Li}_{2}\left(-x\right)+{\rm Li}_{2}\left(-\frac{1}{x}\right)=-\frac{\pi^2}{6}-\frac{1}{2}\ln^2 x~.
 \end{eqnarray}
 A factor $\exp\left(\kappa\pi^2/6\right)$ with $\kappa=(1+2i\nu_0)/(2\chi\left(0,\nu_0\right))$
 has also been absorbed in the normalization constant $S_0$. 
 Taking $\chi\left(0,\nu_0\right)/(1+2i\nu_0)\approx 2.44$ from Eq.\eqref{Mamta}, (initial condition independent)
 constant $S_0$ as unity and defining 
 saturation momentum $Q_s$:  at $x_{10}=1/Q_s$, 
 numerical value of $S$-matrix would be half, $S\left(x_{10}=1/Q_s\right)=1/2$,
 one could now estimate $\lambda_1\approx 7.22$. Therefore, 
 \begin{eqnarray}
 S_{\rm }(x_\perp,Y)=\exp\left(\frac{1+2i\nu_0}{2\chi\left(0,\nu_0\right)} 
 ~{\rm Li}_2\left[-\lambda_1 x_\perp^2Q_s^2(Y)\right]\right)~,   \label{PriyankaChopra2}
 \end{eqnarray}
 with $\lambda_1\approx 7.22$. Eq.\eqref{PriyankaChopra2} is main result of this article. 
 Next we have discussed different limits of Eq.\eqref{PriyankaChopra2}.  \\

 \textbullet ~ In the limit $x_{10}Q_s \ll 1/\lambda_1 \sim 0.14 < 1$  one may retain only the first term in the dilogarithm series, 
 \begin{eqnarray}
 S_{\rm }(x_\perp,Y) \approx \exp\left[-1.48~x_{10}^2 Q^2_s(Y)\right]   ~.
 \end{eqnarray} 
 This is Gaussian in the variable $\tau=x_{10} Q_s(Y)$ in accordance with McLerran-Venugopalan model 
 \cite{McLerran:1993ni,McLerran:1993ka,McLerran:1994vd},
 or Golec-Biernat and M.~Wusthoff model
  \cite{GolecBiernat:1999qd,GolecBiernat:1998js} upto a model dependent variance $\sim 1/3$.  \\ 
 
 
 \textbullet ~In the black disc limit, $x_{10}Q_s \gg \lambda_1\sim 7.2  > 1$, Eq.\eqref{PriyankaChopra} reproduce Levin-Tuchin solution as, 
 \begin{eqnarray}
 S_{\rm }(x_\perp,Y)=\exp\left(-\frac{1+2i\nu_0}{4\chi\left(0,\nu_0\right)}\ln^2\left[x_{10}^2 Q^2_s(Y)\right]\right)   ~,
 \end{eqnarray} \\
 Here we have used the asymptotic expansion of polylogarithms, ${\rm Li}_s\left(z\right)$ in terms of $\ln\left(-z\right)$ and Bernoulli numbers $B_{2k}$ as, 
 \begin{eqnarray}
 {\rm Li}_s\left(z\right) = \sum_{k=0}^{\infty} (-1)^{k}\left(1-2^{1-2k}\right)(2\pi)^{2k}\frac{B_{2k}}{(2k)!}
 \frac{\left[\ln(-z)\right]^{s-2k}}{\Gamma \left(s+1-2k\right)}~.
 \end{eqnarray} \\ 
 
 \noindent In Fig.[\ref{SexyFigure1}] we have plotted 
 dipole amplitude $N(x_{10},Y)$ as function of scaling variable $\tau=x_{10}Q_s(Y)$;
 new solution (Eq.\eqref{AS}) compared with numerical solutions of leading order 
 Balitsky-Kovchegove equation for two sets of rescaled rapidity, $\alpha_s Y=1.2,2.4$, one set of fixed coupling $\alpha_{s}=0.2$. 
 The solution is in better agreement with the numerical solutions of 
 full LO BK equation in a wide kinematic domain inside saturation region.

 \section{conclusion and outlook}
 In this work we have revisited solution of a linearized form of LO BK equation. Unlike the earlier studies here we have adopted 
 transverse width dependent cutoff in order to regulate the dipole integral. We also taken care of all the higher order 
 terms (higher order in cutoff) that have been reasonably neglected before. Later was important in order to make the calculation
 consistent when away from vanishing cutoff. By demanding that dipole integral vanishes in the limit of vanishing transverse 
 separation of the dipole and defining the inverse of saturation momentum being equal to transverse separation of the parent dipole 
 when dipole amplitude is half we derived a general form of solution which reproduce both McLerran-Venugopalan initial conditions 
 (Gaussian in $\tau$) and Levin-Tuchin
 solution  (Gaussian in $\ln \tau$), with $\tau$ being scaling variable, in their appropriate limits. This new solution involving dilogarithm function connects 
 both this limits smoothly and better approximates the numerical estimation of full leading order Balitsky-Kovchegov equation particularly inside 
 saturation
 region. This also implies that linearized LO BK equation contains dynamics of dipole nucleus interaction throughout a 
 wide kinematic domain of saturation. It would be interestingly to see how this solution modifies for the  running 
 couplings improved or next to leading order BK equations, how it preserves the inherent conformal symmetry of the kernel,
 or to what extent it receives corrections from quadratic nonlinear term (in $S$-matrix) present in the Balitsky-Kovchegov equation.

 \begin{acknowledgments}
 We are indebted to Yuri Kovchegov for many valuable suggestions and comments since inception of this work. We also 
 thank Rafi Alam, Trambak Bhattecharya, Haider H. Jafri and Manjari Sharma for valuable discussions and help.

 \end{acknowledgments}

  \vspace{2cm}

  \subsection*{Appendix: Calculation of $I_{\rm dip}$}
  Here we detailed derivation of Eq.\eqref{Moutusi}. Substituting Eq. \eqref{Alia}  into Eq. \eqref{Madhuri} we obtain, 
   \begin{eqnarray}
   I_{\rm dip} &=& 2\pi x_{10}^2\int_{0}^{\infty} dk k J(x_{10}k) 
                    \left( \ln \frac{2}{k\rho} - \gamma_{E}+ \sum_{m=1}^{\infty} (-1)^{m+1}{\cal C}_{2m}k^{2m}\rho^{2m}  \right)^2  \nn \\
               &=& I_{0}+I_{1}+I_{2}
  \end{eqnarray} 
  where, 
 \begin{eqnarray}
  I_0 &=& 2\pi x_{10}^2\int_{0}^{\infty} dk k J(x_{10}k) \left( \ln \frac{2}{k\rho} - \gamma_{E}\right)^2  \nn \\ 
  I_1 &=& 2\pi x_{10}^2\int_{0}^{\infty} dk k J(x_{10}k) \left( \sum_{m=1}^{\infty} (-1)^{m+1}{\cal C}_{2m}k^{2m}\rho^{2m} \right)^2 \nn \\ 
  I_2 &=& 4\pi x_{10}^2\int_{0}^{\infty} dk k J(x_{10}k) \left( \ln \frac{2}{k\rho} 
  - \gamma_{E}  \right) \left( \sum_{m=1}^{\infty} (-1)^{m+1}{\cal C}_{2m}k^{2m}\rho^{2m} \right) \label{I_2}
  \end{eqnarray} 
   with,  
  \begin{eqnarray}
  {\cal C}_{2m}=\frac{1}{2m}\frac{1}{2^{2m}(m!)^2}
  \end{eqnarray}
 Using, 
 \begin{eqnarray}
   \int_{0}^{\infty} dk~k~k^{p}~J_{0}(x_{10}k) = 0   \label{Sunny}
  \end{eqnarray}
  where $x_{10} > 0$ and $p$ is either zero or positive even integer,  
 \begin{eqnarray}
  \int_{0}^{\infty} dk~k~J_{0}(x_{10}k)~\ln k
  =\lim_{\epsilon\rightarrow 0} \frac{\partial}{\partial \epsilon} \int_{0}^{\infty} dk~k^{1+\epsilon}~J_{0}(x_{10}k) = -\frac{1}{x_{10}^2}
\label{Madhabi1}
  \end{eqnarray}  
 and, 
 \begin{eqnarray}
 \int_{0}^{\infty} dk~k~J_{0}(x_{10}k)~\ln^2 k
 =\lim_{\epsilon\rightarrow 0} \frac{\partial^2}{\partial \epsilon^2} \int_{0}^{\infty} dk~k^{1+\epsilon}~J_{0}(x_{10}k)
 =\frac{2}{x_{10}^2}\left(\ln\frac{x_{10}}{2}+\gamma_E\right)
 \label{Madhabi2}
 \end{eqnarray} 
 all of which follows from, 
 \begin{eqnarray}
 \int_{0}^{\infty} dk k^{\lambda-1} J_0\left(kx\right)
 =2^{\lambda-1}x^{-\lambda}\frac{\Gamma\left(\lambda/2\right)}{\Gamma\left(1-\frac{\lambda}{2}\right)} ~.  \label{hot}
 \end{eqnarray} 
 Using Eq.\eqref{Madhabi1} and Eq.\eqref{Madhabi2} integral $I_0$ can be written as, 
 \begin{eqnarray}
   I_0 &=& 2\pi x_{10}^2\int_{0}^{\infty} dk k J(x_{10}k) 
                    \left( \ln \frac{2}{k\rho} - \gamma_{E}\right)^2  \nn \\
               &=& 2\pi \ln\left(\frac{x_{10}^2}{\rho^2}\right)
 \end{eqnarray} 
 Eq.[\ref{Sunny}] ensures that $I_{1}$ vanishes for $x_{10}>0$, 
 \begin{eqnarray}
   I_1 &=& 2\pi x_{10}^2\int_{0}^{\infty} dk k J(x_{10}k) 
                    \left( \sum_{m=1}^{\infty} (-1)^{m+1}{\cal C}_{2m}k^{2m}\rho^{2m} \right)^2  = 0
  \end{eqnarray}  
 and terms containing $\gamma_E$ and $\ln 2$ in the integral $I_2$ in Eq.\eqref{I_2} will vanish as well, 
  \begin{eqnarray}
   I_2 &=& 4\pi x_{10}^2\int_{0}^{\infty} dk k J(x_{10}k) 
   \left( \ln \frac{2}{k\rho} - \gamma_{E}  \right) \left( \sum_{m=1}^{\infty} (-1)^{m+1}{\cal C}_{2m}k^{2m}\rho^{2m} \right)  \nn \\
       &=& -4\pi x_{10}^2\sum_{m=1}^{\infty} (-1)^{m+1}{\cal C}_{2m} \int_{0}^{\infty}~dk~k~J(x_{10}k)~\ln (k\rho)~(k\rho)^{2m}  \nn \\
       &=& -4\pi x_{10}^2\sum_{m=1}^{\infty} (-1)^{m+1}{\cal C}_{2m} \lim_{\epsilon\rightarrow 0}\frac{\partial}{\partial \epsilon}
                    \int_{0}^{\infty}~dk~k~J(x_{10}k)~(k\rho)^{2m+\epsilon} ~. 
  \end{eqnarray}  
  Eq.\eqref{hot} could be use to evaluate the integral,  
  \begin{eqnarray}
     I_2 &=& -4\pi \frac{x_{10}^2}{\rho^2}\sum_{m=1}^{\infty} (-1)^{m+1}{\cal C}_{2m} 
               \lim_{\epsilon\rightarrow 0} \frac{\partial}{\partial \epsilon}
               ~ 2^{2m+1+\epsilon}~\left(\frac{x_{10}}{\rho}\right)^{-2m-2-\epsilon}
               ~\frac{\Gamma(m+1+\epsilon/2)}{\Gamma\left(-m-\epsilon/2\right)}  \nn \\                
    &=& -4\pi \sum_{m=1}^{\infty} (-1)^{m+1}~{\cal C}_{2m}~ 2^{2m+1}~\left(\frac{\rho}{x_{10}}\right)^{2m}
                ~\lim_{\epsilon\rightarrow 0} \frac{\partial}{\partial \epsilon}
               ~ 2^{\epsilon}~\left(\frac{x_{10}}{\rho}\right)^{-\epsilon}~\frac{\Gamma(1+\epsilon/2)}{\Gamma\left(-\epsilon/2\right)} ~
               \prod_{i=1}^{m}\left(i+\frac{\epsilon}{2}\right)^2 \nn \\
       &=& -4\pi \sum_{m=1}^{\infty} (-1)^{m+1}{\cal C}_{2m} 2^{2m+1}~(-1)^{m}~(m!)^2~\left(\frac{\rho}{x_{10}}\right)^{2m}
                ~\left(-\frac{1}{2}\right) \nn \\          
     &=& -4\pi \sum_{m=1}^{\infty}~{\cal C}_{2m} ~2^{2m}~(m!)^2~\left(\frac{\rho}{x_{10}}\right)^{2m}
                \nn     \\ 
      &=& -2\pi \sum_{m=1}^{\infty}~\frac{1}{m}~\left(\frac{\rho^2}{x_{10}^2}\right)^{m}
                \nn     \\           
     &=&   -2\pi ~ {\rm Li}_{1}\left(\frac{\rho^2}{x_{10}^2}\right)     
 \end{eqnarray}  
 Finally the dipole integral is, 
  \begin{eqnarray}
   I_{\rm dip}
   &=& I_{0}+I_{1}+I_{2}  \nn \\ 
   &=&  2\pi \ln\left(\frac{x_{10}^2}{\rho^2}\right)-2\pi ~ {\rm Li}_{1}\left(\frac{\rho^2}{x_{10}^2}\right)  \nn \\
   &=&  2\pi \ln\left(\lambda_1 x_{10}^2Q_s^2+\lambda_2\right) 
  -2\pi {\rm Li}_{1}\left(\frac{1}{\lambda_1 x_{10}^2Q_s^2+\lambda_2}\right)~, 
  \end{eqnarray}  
  where we have substituted $\rho$ by $Q_s$ and $x_{10}$ using Eq.\eqref{AngelinaJolie4} in the last line. 

\end{document}